\documentclass[%
 reprint,%
 amssymb, amsmath,%
 aip,cha%
]{revtex4-1}
\usepackage[columnwise]{lineno}

\usepackage{docs}%
\usepackage{bm}%
\usepackage[colorlinks=true,linkcolor=blue]{hyperref}%
\expandafter\ifx\csname package@font\endcsname\relax\else
 \expandafter\expandafter
 \expandafter\usepackage
 \expandafter\expandafter
 \expandafter{\csname package@font\endcsname}%
\fi
\hypersetup{
	colorlinks=true,
	linkcolor=blue,
	filecolor=magenta,}
\usepackage{graphicx}
\usepackage{lipsum}

\begin{document}

\title{Study of magnetic interface and its effect in Fe/NiFe bilayers of alternating order}

\author{Sagarika Nayak}%
\affiliation{Laboratory for Nanomagnetism and Magnetic Materials (LNMM), School of Physical Sciences, National Institute of Science Education and Research (NISER), HBNI, P.O.- Jatni, 752050, India}%
\author{Sudhansu Sekhar Das}%
\affiliation{Laboratory for Nanomagnetism and Magnetic Materials (LNMM), School of Physical Sciences, National Institute of Science Education and Research (NISER), HBNI, P.O.- Jatni, 752050, India}%
\author{Braj Bhusan Singh}%
\affiliation{Laboratory for Nanomagnetism and Magnetic Materials (LNMM), School of Physical Sciences, National Institute of Science Education and Research (NISER), HBNI, P.O.- Jatni, 752050, India}%
\author{Timothy R. Charlton}%
\affiliation{Oak Ridge National Laboratory, 1 Bethel Valley Rd, Oak Ridge, TN 37830, United States}%
\author{Christy J. Kinane}%
\affiliation{ISIS, Harwell Science and Innovation Campus, Science and Technology Facilities Council, Rutherford Appleton Laboratory, Didcot, Oxon OX11 0QX, United Kingdom}%
\author{Subhankar Bedanta}%
\email{sbedanta@niser.ac.in}
\affiliation{Laboratory for Nanomagnetism and Magnetic Materials (LNMM), School of Physical Sciences, National Institute of Science Education and Research (NISER), HBNI, P.O.- Jatni, 752050, India}%
\date{April 2019}%


\begin{abstract}
We present a comprehensive study on the magnetization reversal in Fe/NiFe bilayer system by alternating the order of the magnetic layers. All the samples show growth-induced uniaxial magnetic anisotropy due to oblique angle deposition technique. Strong interfacial exchange coupling between the Fe and NiFe layers leads to the single-phase hysteresis loops in the bilayer system. The strength of coupling being dependent on the interface changes upon alternating the order of magnetic layers. The magnetic parameters such as coercivity $H_{C}$, and anisotropy field $H_{K}$ become almost doubled when NiFe layer is grown over the Fe layers. This enhancement in the magnetic parameters is primarily dependent on the increase of the thickness and magnetic moment of Fe-NiFe interfacial layer as revealed from the polarized neutron reflectivity (PNR) data of the bilayer samples. The difference in the thickness and magnetization of the Fe-NiFe interfacial layer indicates the modification of the microstructure by alternating the order of the magnetic layers of the bilayers. The interfacial magnetic moment increased by almost 18 $ \% $ when NiFe layer is grown over the Fe layer. In spite of the different values of anisotropy fields and modified interfacial exchange coupling, the Gilbert damping constant values of the ferromagnetic bilayers remain similar to single NiFe layer.
 
\end{abstract}

\maketitle
In an exchange-coupled soft/hard bilayers, one can find high energy product $(BH)_{max}$ value as the soft magnetic layer provides high saturation magnetization $M_{S}$ and the hard one provides intermediate coercivity $H_{C}$~\cite{Wei2013}. This soft-hard combination of the magnetic bilayers provides an excellent research opportunity not just for their potential  application in the field of permanent magnets~\cite{Wei2013} but also for the sake of fundamental understanding of various magnetization reversal processes. Hard magnetic layer gives large $H_{C}$ due to its high magnetic anisotropy which is not desired in the application of the write head. Further, the switching field of the hard layer can be reduced by fabricating soft/hard magnetic bilayers which fulfills the requirement of write-head and simultaneously provide excellent temperature stability~\cite{Wang2009,Alexandrakis2010}. The interface plays a very important role in tuning the $H_{C}$ of the bilayers by modifying the interfacial exchange coupling. In literature, several techniques have been employed for modifications of the interfacial layer. Different deposition techniques and subsequent post-deposition annealing at different temperatures have been widely employed for modification of the interface~\cite{Wang2009, Alexandrakis2010, Crew2001}. Varying the thickness of the soft layer and using the materials of different crystallographic structures have also been considered to study the role of interfacial exchange coupling in a soft/hard bilayer system ~\cite{ Wei2013, Wang2009,Wang2013}. The interfacial exchange coupling between the hard and soft layers can be enhanced by interdiffusion~\cite{Alexandrakis2009}. Although there are several reports on various techniques for modification of the interface, there is still a continuous focus to understand the interface properties~\cite{Behera2017, Tang2017}. 

In order to account the role of interdiffusion on the magnetic properties, several experiments and simulations have been performed. Conversion Electron  M$\ddot{o}$ssbauer Spectroscopy has been used to find the presence of interdiffusion in hard (FePt)/soft (Fe or Co/Fe) bilayers~\cite{Reddy2012}. The presence of a graded interface has been observed in SmCo/Fe system from Synchrotron x-ray scattering and electron microscopy elemental mapping measurements~\cite{Jiang2004}. Transmission electron microscopy and magnetic measurements show an enhanced epitaxy in the postannealed Co/CoPt system~\cite{Kim2000}. Depth and element resolved x-ray resonant magnetic scattering (XRMS) measurements on SmCo/Fe bilayer show the presence of diffused Co-atoms from Sm-Co layer in Fe magnetic layer~\cite{Choi2007}. Despite several experimental techniques for studying the magnetic properties of the individual layers, a quantitative knowledge about the interface of a layered system always remains challenging due to the very complex nature of the interface~\cite{Liu2011}. In this context, polarized neutron reflectivity (PNR) is a very promising tool for a quantitative structural and magnetic information about the interface.

In addition to the above study, materials with lower Gilbert damping constant $\alpha$ are being studied extensively for their application in spin-transfer torque-based oscillators~\cite{Conca2013}, and also in spintronics devices~\cite{Pal2011}. Intrinsic Gilbert damping in materials has its origin on the spin-orbit coupling~\cite{Hickey2009}. Extrinsic contributions can enhance this damping. Several deposition methodologies such as different oblique angle of deposition\cite{Phuoc2009}, deposition pressure~\cite{Xu2012} etc. have been employed for tuning of the damping constant. It is desired to fabricate hard/soft magnetic bilayers where anisotropy gets enhanced keeping the damping value of the same order as that of the reference soft layer. 

In the present paper, we report tuning  of the interfacial exchange coupling by alternating the order of magnetic layers in the hard/soft Fe/NiFe bilayers. We show that by alternating the order of layers the interface changes which results in tuning of the magnetic properties of the bilayers. In order to quantify the interface thickness and moment, we have performed polarized neutron reflectometry on the bilayer samples. We also made a comparative study on the damping constants of the samples through FMR analysis.
\section{Experimental details:}
All the samples are deposited by combination of dc magnetron sputtering and e-beam deposition in a high vacuum chamber on naturally oxidized Si (100) substrate. The base pressure was $\sim$ 6$\times$10$^{-8}$ mbar. Prior to the deposition, the substrates were annealed for a period of 2 hr at 150 $^{\circ}$C. The samples were prepared on to the Si-substrates kept at 150 $^{\circ}$C at an Ar pressure of $\sim$ 5$\times$10$^{-3}$ mbar. A capping layer of Au (3nm) was further deposited by e-beam evaporation to protect the samples from oxidation.  The rate of deposition of Fe, NiFe and Au were kept at 0.22, 0.17 and 0.1 \AA{}/sec, respectively. Table 1  shows the list of sample nomenclature, and structure. 

\begin{table}[hbt!]
\small
  \caption{\ Details of sample name and structure for all the samples.}
  \label{tbl:example}
  \begin{tabular*}{0.48\textwidth}{@{\extracolsep{\fill}}ll}
    \hline
    Sample name&Sample structure \\
    \hline
    S1&Si (100)/NiFe (10 nm)/Au (3 nm) \\
	S2&Si (100)/Fe (5 nm)/Au (3 nm) \\
	S3&Si (100)/Fe (5 nm)/NiFe (10 nm)/Au (3 nm) \\
	S4&Si (100)/NiFe (10 nm)/Fe (5 nm)/Au (3 nm) \\
    \hline
  \end{tabular*}
\end{table}
We performed x-ray reflectivity (XRR) measurements to evaluate the thickness, density and roughness of each individual layers by using x-ray diffractometer from Rigaku with CuK$_{\alpha}$ radiation ($\lambda$ = 0.154 nm). We have performed PNR measurements at room temperature at POLREF neutron reflectometer, at Rutherford Appleton Laboratory, UK. In the PNR measurements, magnetic field was applied along the easy axis (EA) and experiments were performed at saturation and near to coercive field of the bilayer samples. POLREF is a white beam instrument and we have used a pulse of length 2-15 \AA{}'s with several varying angles. We plotted the experimental data of reflectivity vs perpendicular scattering vector $Q_{Z}$ = 4$\pi$sin$\theta$/$\lambda$ where $Q_{Z}$ is the component of momentum transfer perpendicular to the sample surface, thus, giving sample’s layer-by-layer information~\cite{Malik2019,Mallik2018a}. We always applied positive saturation field and then reverse the field to the measurement fields. The guiding field was -1 mT. Neutron reflectivity can be spin flipped or non-spin flipped. We measured two non-spin flipped scattering cross sections namely R$^{++}$ and R$^{--}$~\cite{Malik2019,Mallik2018a}. In R$^{++}$, the first + sign is for the incident neutron with up-spin polarization and the second + sign is for the reflected neutron with up-spin polarization. Similarly, we can explain R$^{--}$ (down-down). The XRR and PNR data were fitted using GenX software~\cite{GenX}.  We have performed longitudinal magneto-optic Kerr effect based microscopy to simultaneously measure hysteresis loops and image the magnetic domains. Magnetic dynamic properties were studied using ferromagnetic resonance (FMR) setup manufactured by Nano Osc.
\section{Results and discussion:}
\subsection{PNR analysis}
We have evaluated the quantitative structural information such as density, roughness and thickness of the samples from x-ray reflectivity (XRR) measurement (data are not shown). The layer thickness and roughness obtained from XRR and PNR measurements are similar for the bilayer samples.

\begin{table*}[hbt!]
\small
  \caption{\ Structural and magnetic parameters obtained after fitting the PNR experimental data using GenX software for sample S3.}
  \label{tbl:example}
  \footnotesize
  \begin{tabular*}{\textwidth}{@{\extracolsep{\fill}}lllll}
    \hline
    Layer description&thickness (nm)&roughness (nm)&Magnetic moment ($\mu_{B}$/atom) at -50 mT&Magnetic moment ($\mu_{B}$/atom) at -4 mT \\ \\
    \hline
    Au & 3.79 & 1.99 & -- & --  \\
	NiFe-Au & 2.99 & 0.94 & -0.10 & -0.003  \\
	NiFe & 8.39 & 1.09 & -0.79 & 0.78  \\
	Fe-NiFe & 2.31 & 1.20 & -0.90 & 0.8  \\	
	Fe & 2.48 & 1.56 & -1.57 & 1.57  \\
	Si$O_{2}$-Fe & 1.99 & 0.99 & -1.00 & 0.99  \\
    \hline
  \end{tabular*}
\end{table*}
\begin{table*}[hbt!]
\small
  \caption{\ Structural and magnetic parameters obtained after fitting the PNR experimental data using GenX software for sample S4.}
  \label{tbl:example}
  \footnotesize
  \begin{tabular*}{\textwidth}{@{\extracolsep{\fill}}lllll}
    \hline
    Layer description&thickness (nm)&roughness (nm)&Magnetic moment ($\mu_{B}$/atom) at -50 mT&Magnetic moment ($\mu_{B}$/atom) at -1.2 mT \\ \\
    \hline
    Au & 3.00 & 1.37 & -- & --  \\
	Fe-Au & 2.66 & 1.19 & -0.52 & 0.20  \\
	Fe & 3.24 & 0.85 & -1.26 & 1.17  \\
	NiFe-Fe & 1.79 & 1.10 & -0.76 & 0.76  \\	
	NiFe & 8.99 & 1.29 & -0.75 & 0.75  \\
	Si$O_{2}$-NiFe & 1.66 & 0.9 & -0.75 & 0.75  \\
    \hline
  \end{tabular*}
\end{table*}
In order to get quantitative information from the layers and interfaces in the sample stack, we have performed polarized neutron reflectivity (PNR) measurement on the samples. PNR has been proven to be an ideal technique for providing layer-by-layer magnetization profile in a multilayer stack. We have fitted the PNR data by considering different interface models to find the best figure of merit (FOM). Considering all other interface models other than the three interface model, we found less value of FOM and the fitting is not good. We found that best FOM is achieved by considering a three interface model in samples S3 and S4. The interfaces are named as NiFe-Au, Fe-NiFe and Si$O_{2}$-Fe for sample S3. Similarly the interfaces are named as Fe-Au, NiFe-Fe and Si$O_{2}$-NiFe for sample S4. The interfaces taken to fit the neutron reflectivity data are shown in fig. 1.  FOMs of 4.60$\times$10$^{-02}$ and 5.04$\times$10$^{-02}$ are found in samples S3 and S4. Here, we have used LOG type of FOM. Using this type of FOM, we fitted the data more easily and robustly. LOG type of FOM takes into account the average of the difference between the logarithms of the data and the simulation. Structural and magnetic parameters, obtained from PNR fit, are shown in tables 2 and 3 for samples S3 and S4, respectively. The magnetic moment of Fe and NiFe obtained from the PNR data of the sample S3 are 1.57 $\mu_{B}$/atom and 0.7 $\mu_{B}$/atom, respectively. The observed deviation in the magnetic moment of Fe and NiFe from their bulk value is due to the transfer of magnetic moment to the interface caused by interdiffusion. Similarly, the Fe-NiFe interface of sample S3 has a magnetic moment of 0.90  $\mu_{B}$/atom which is intermediate between Fe and NiFe layers, and has a thickness of 2.3 nm. The Si$O_{2}$-Fe interface has lesser magnetic moment of 1.00 $\mu_{B}$/atom than Fe itself due to interdiffusion in sample S3~\cite{Singh2007}. Interface roughness of the order of  1 nm might also be a reason for lesser magnetic moment at Fe-Si$O_{2}$ interface~\cite{Singh2007}. In contrast to the Si$O_{2}$-Fe interface, the NiFe-Au interface in the sample S3 has a relatively smaller values of magnetic moment (0.10 $\mu_{B}$/atom) and thickness(2.99 nm), indicating high amount of  interdiffusion. Thus, a dead layer is formed at the NiFe-Au interface of sample S3. Our XRR data also suggests different rougnness values of Fe and NiFe which are in direct contact with Si$O_{2}$ and Au layers, respectively. Thus, we can conclude that interface roughness might be a reason for the different values of magnetic moment at Si$O_{2}$-Fe and NiFe-Au interfaces than the parent layers in sample S3. Interdiffusion and/or alloying at the interfaces are the result of high temperature (150 {$^\circ$}C) deposition of the studied films. The formation of a magnetic dead layer is also reported in the case of Fe/Ge system when Fe is grown on Ge at 150 {$^\circ$}C ~\cite{Singh2007}. In contrast to the sample S3, the magnetic moment of the Fe layer, NiFe layer and the NiFe-Fe interfacial layer in the sample S4, are 1.26 $\mu_{B}$/atom, 0.75 $\mu_{B}$/atom, and 0.76 $\mu_{B}$/atom, respectively. This indicates that the sample S3 has relatively higher magnetic moment values of its constituent layers Fe, NiFe and Fe-NiFe interface as compared to that of the S4. This is further confirmed from the SQUID data of the samples where the S3 has a higher saturation magnetization value (762 emu.$cc^{-1}$) in comparison to that of S4 (636 emu.$cc^{-1}$).
\begin{figure}[hbt!]
\centering
  \includegraphics[height=5cm]{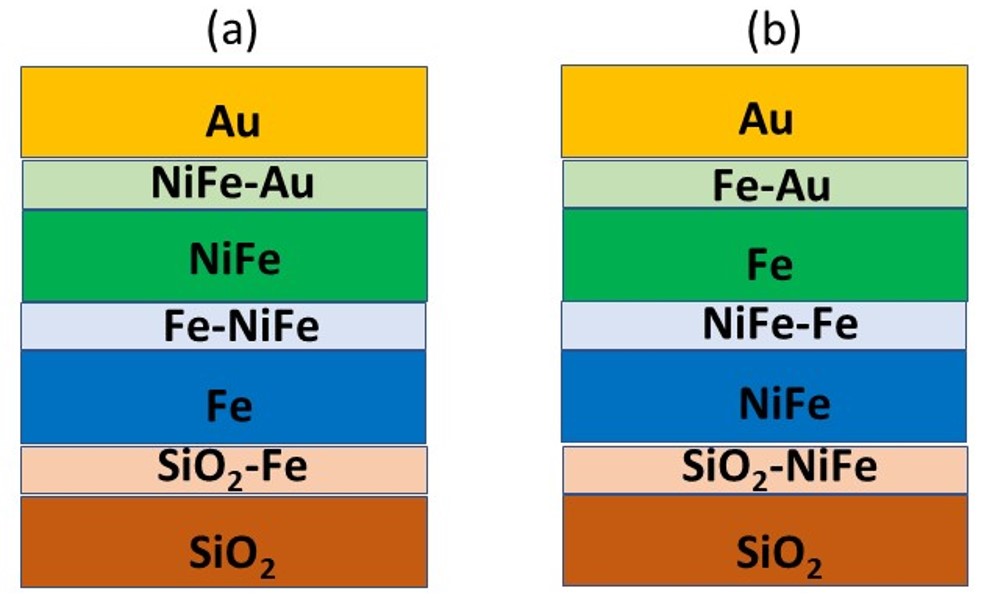}
  \caption{Schematic of all the interfaces and thin film layers in samples (a) S3 and (b) S4.}
  \label{fig:fig1}
\end{figure}
\begin{figure*}[hbt!]
	\centering
	\includegraphics[height=6cm]{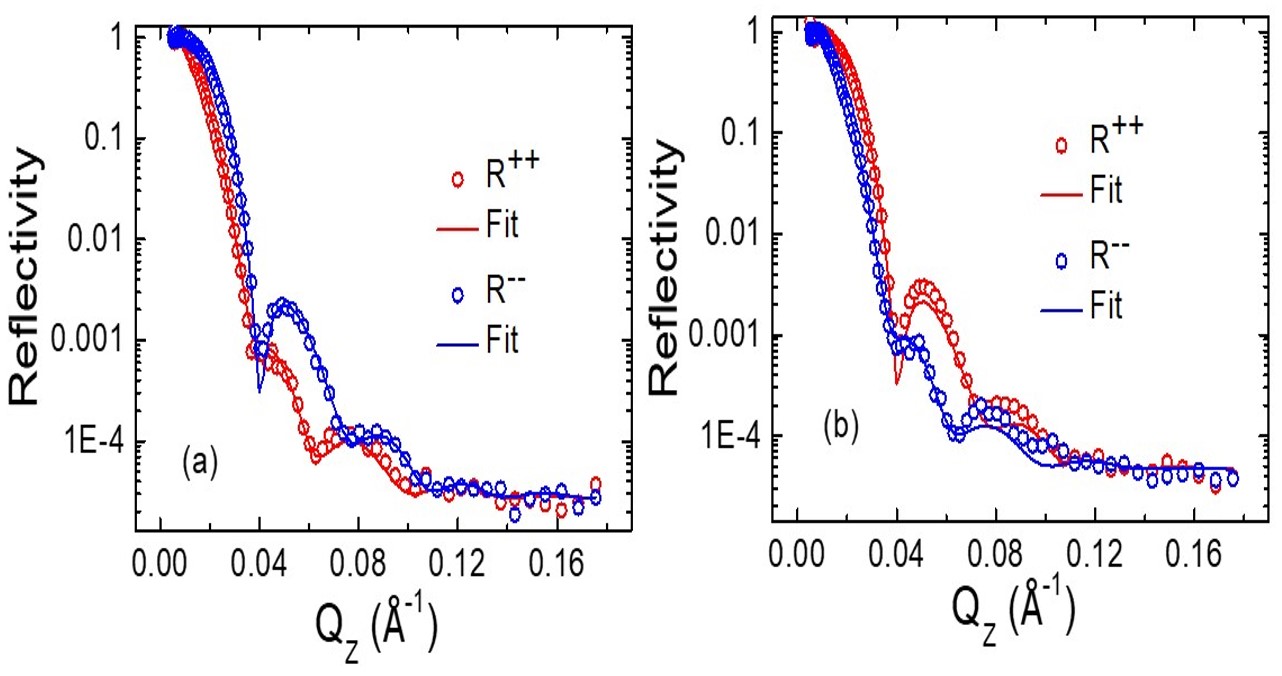}
	\caption{Polarized neutron reflectivity (PNR) data for sample S3 at room temperature with saturation magnetic field of -50 mT (a) and -4 mT of magnetic field which is near to coercivity (b) are applied along EA. The open circles are the experimental data points and the solid lines are fitted data for the non-spin flip (NSF) reflectivities $R^{++}$ (red colour), $R^{--}$ (blue colour), respectively.}
	\label{fig:fig8}
\end{figure*}
\begin{figure*}[hbt!]
	\centering
	\includegraphics[height=6cm]{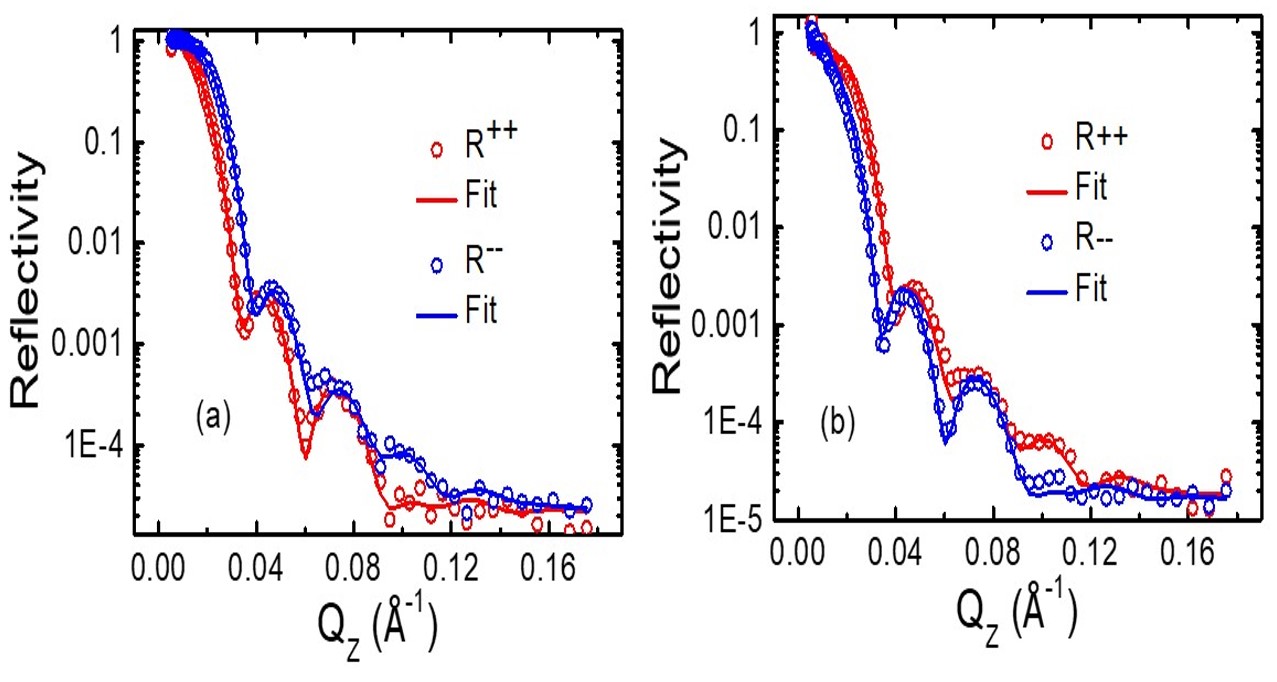}
	\caption{Polarized neutron reflectivity (PNR) data for sample S4 at room temperature measured at saturation magnetic field of -50 mT (a) and -1.2 mT of magnetic field near to coercivity (b), along EA. The open circles are the experimental data points and the solid lines are fitted data for the non-spin flip (NSF) reflectivities $R^{++}$ (red colour), $R^{--}$ (blue colour), respectively. }
	\label{fig:fig9}
\end{figure*}
We observed that all magnetic layers including  Si$O_{2}$-Fe interface are reversed completely at -4 mT of magnetic field in sample S3.  88 $\%$ of Fe-NiFe interface magnetic moments are reversed from positive  saturation state at -4 mT field in sample S3. Further, 92 $\%$ of magnetic moments of Fe have reversed their direction near to coercive field (-1.2 mT) from positive saturation state in sample S4. Again, 38 $\%$ of the magnetic moment at the Fe-Au interface has reversed direction in sample S4 whereas all other layers has reversed completely. Further, the thicknesses of all the interdiffused interface layers of sample S3 are higher than that of sample S4. Thus, larger interdiffusion might be a reason for the difference in magnetic properties of samples S3 and S4. We found from tables 2 and 3 that the roughness of Fe in sample S3 is higher than sample S4 whereas the magnetic moment of Fe is higher in sample S3. Similarly, Fe-NiFe interface of sample S3 has higher roughness and magnetic moment than sample S4. The thickness of NiFe magnetic layer is higher in sample S4, and hence, higher roughness in comparison to sample S3.
We found high values of interdiffusion layer thickness and magnetic moment at Fe-NiFe interface in sample S3 as compared to sample S4. Also, a dead layer is created at the NiFe-Au interface in sample S3 whereas no dead layer is formed in the Fe-Au interface in sample S4. The presence of high exchange coupling may be a possible reason for the higher value of coercivity and anisotropy field $H_{K}$ in sample S3 than S4 (see table 4). 

\begin{figure*}[hbt!]
	\centering
	\includegraphics[height=6cm]{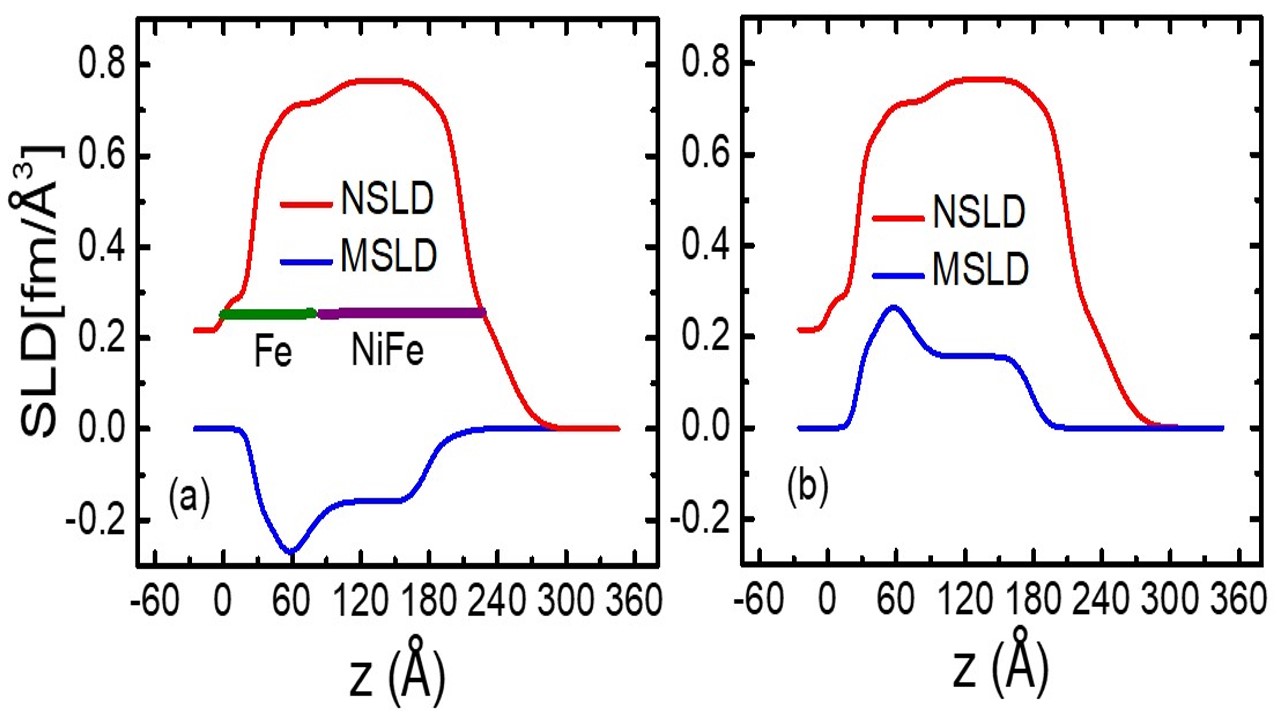}
	\caption{Nuclear and magnetic scattering length densities (NSLD and MSLD) vs layer thickness (z) of the sample S3 at saturation field of -50 mT (a) and near to $H_{C}$ at -4 mT (b).}
	\label{fig:fig9}
\end{figure*}
\begin{figure*}[hbt!]
	\centering
	\includegraphics[height=6cm]{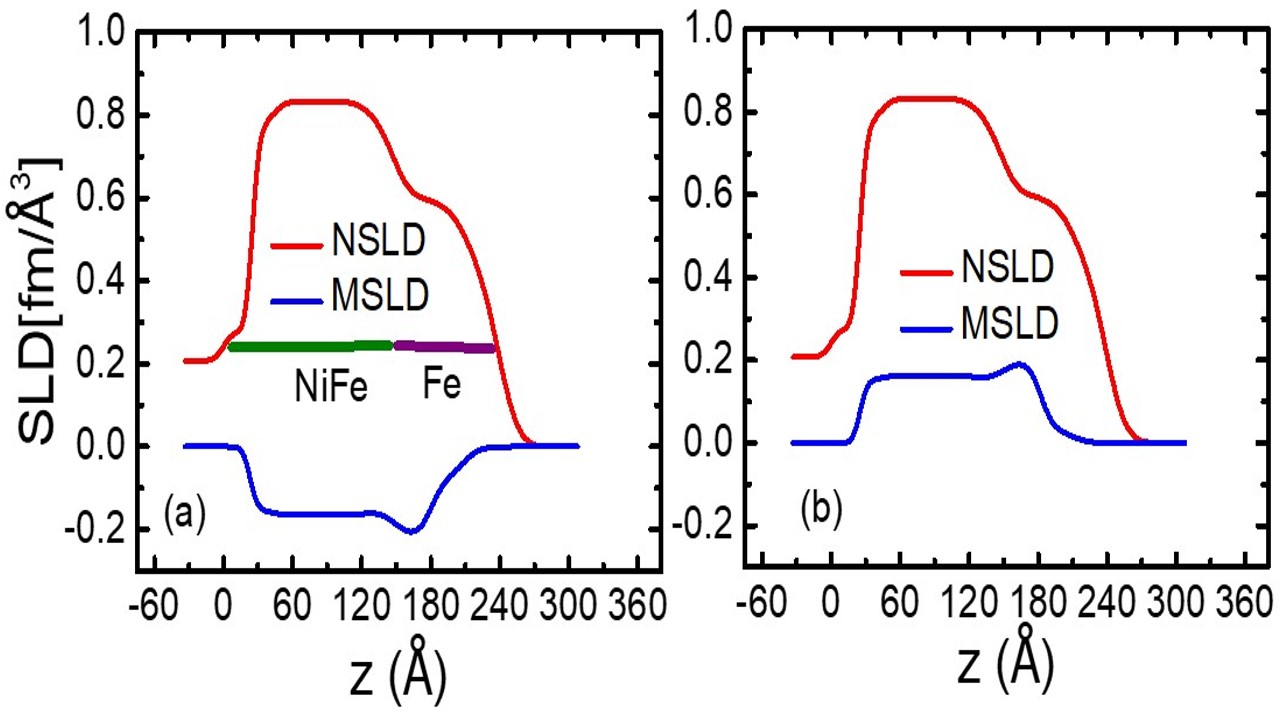}
	\caption{Nuclear and magnetic scattering length densities (NSLD and MSLD) vs layer thickness (z) of the sample S4 at saturation field of -50 mT (a) and near to $H_{C}$ at -1.2 mT (b).}
	\label{fig:fig9}
\end{figure*}
The nuclear scattering length density (NSLD) is found to be of 0.2 fm/$\text{\normalfont\AA}^{3}$ near to Si$O_{2}$-Fe interface whereas zero NSLD is found above NiFe-Au interface in sample S3. However, we found zero magnetic scattering length density (MSLD) near to Si$O_{2}$-Fe and NiFe-Au interfaces of sample S3. Similar trends of NSLD and MSLD is found near to the interfaces Si$O_{2}$-NiFe and Fe-Au. Also, we found the similar trend of the NSLD and MSLD profiles near to saturation and $H_{C}$ field values for the samples S3 and S4. Comparing the SLD profiles of samples S3 and S4, we found a sharp drop in NSLD for Fe magnetic layer of sample S4 whereas SLD is almost constant for Fe and NiFe layers of sample S3. Also, we found the change in sign of the MSLD's for the samples S3 and S4 near to the coercive field $H_{C}$ and this is due to magnetic field history. We can not say that depolarisation is responsible for the sign chnage of MSLD because the PNR measurement fields (shown in the figures 6 (c) and (d) with green coloured square symbols) and guide field are along the same direction.

We can calculate the MSLD from the $M_{S}$ obtained from SQUID using the relation MSLD=C.$M_{S}$, where C=2.853 $\times$ $10^{-9}$ ($\text{\normalfont\AA}^{-2}$).($cm^{3}$/emu). We found MSLD of 0.22 and 0.18 fm/$\text{\normalfont\AA}^{3}$ for the samples S3 and S4. These MSLD values obtained from SQUID match well with the values found from PNR (see figures 5 and 6).
\subsection{Kerr microscopy and magnetometry analysis}
\begin{table}[hbt!]
\center
	\caption {$H_{C}$ along EA and HA and $H_{K}$ for all the samples.}
	\footnotesize
	\begin{tabular}{@{}lllll}
		\hline
		Sample name&$H_{C}$ (EA) (mT)&$H_{C}$ (HA) (mT)&$H_{K}$ (mT) \\       
		\hline
		S1&0.80&0.38&4.28\\
		S2&0.74&0.32&2.44\\
		S3&5.10&1.47&7.10\\
		S4&1.45&0.79&4.00\\
		\hline
	\end{tabular}\\
\end{table}
\begin{figure*}[hbt!]
	\centering
	\includegraphics[height=10cm]{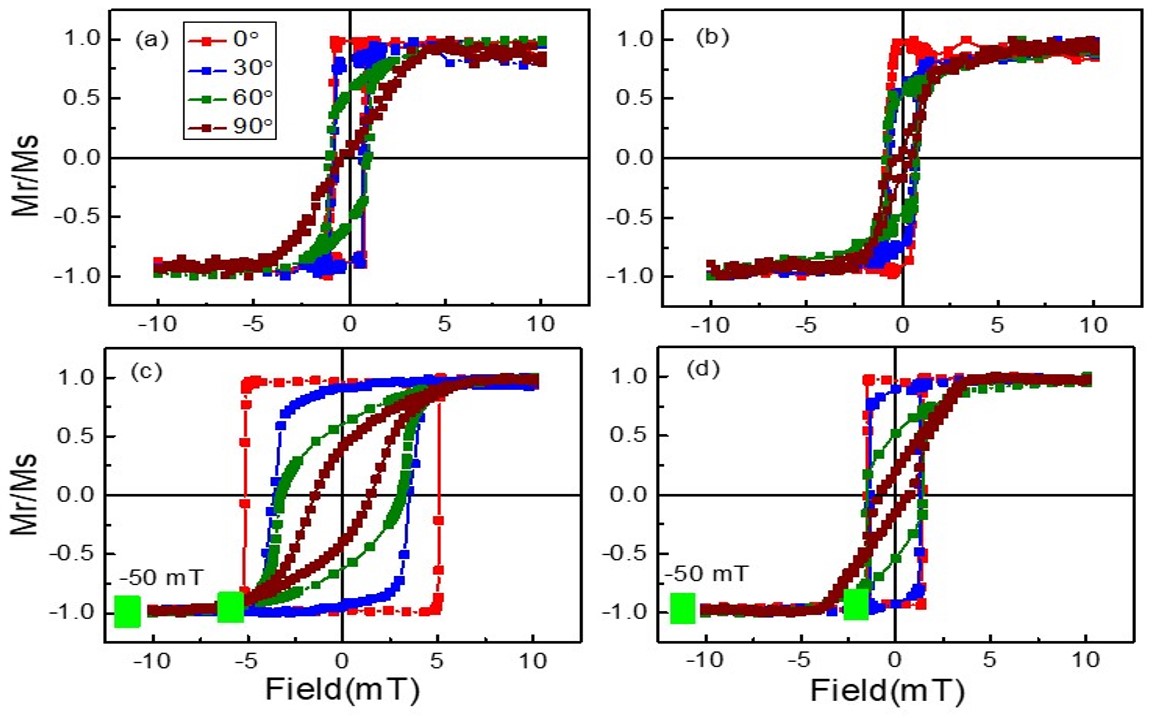}
	\caption{(a)-(d): Hysteresis loops measured by LMOKE at room temperature along $\phi$ = 0$^{\circ}$, 30$^{\circ}$, 60$^{\circ}$, and 90$^{\circ}$ for samples S1-S4.}
	\label{fig:fig2}
\end{figure*}
Hysteresis loops were measured using longitudinal magneto optic Kerr effect (LMOKE) based magnetometry at room temperature along $\phi$ = 0$^{\circ}$ (EA), 30$^{\circ}$, 60$^{\circ}$, 90$^{\circ}$ w.r.t. EA for all the samples, which are shown in figure 6. We observed square-shaped loops along EA and s-shaped loops along HA for all the samples. This indicates, magnetization reversal is occuring via domain wall motion along EA and coherent rotation along HA. From the hysteresis loops, it is also concluded that the samples exhibit uniaxial magnetic anisotropy due to oblique angle of deposition. It is reported in the literature
\begin{figure*}[hbt!]
	\centering
	\includegraphics[height=9cm]{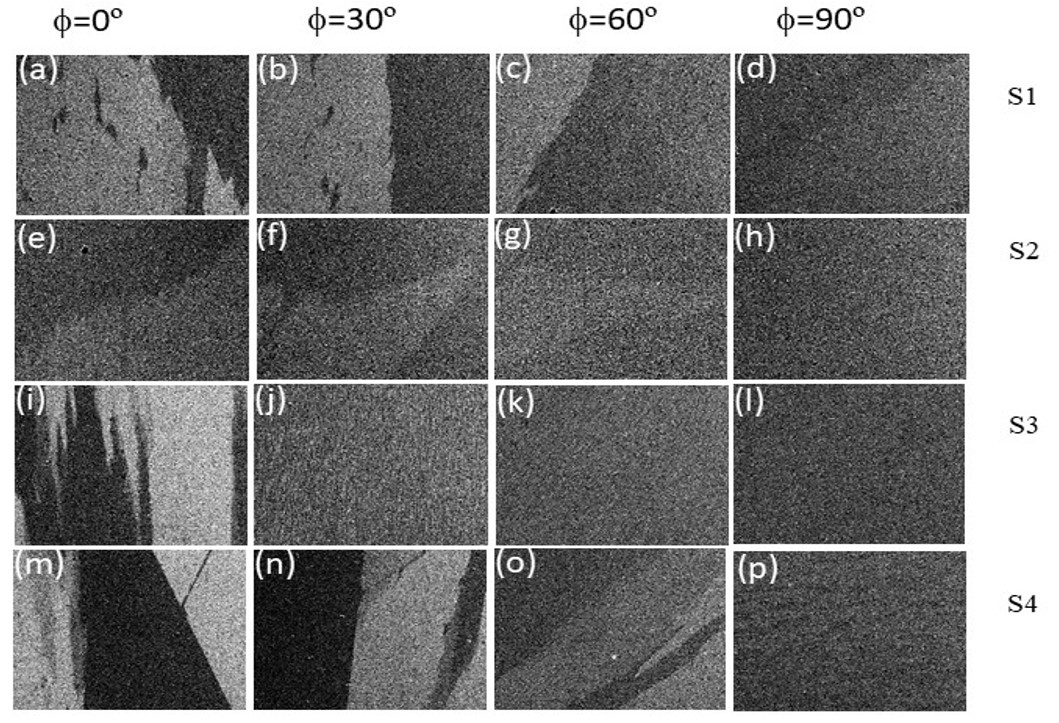}
	\caption{Magnetic domain images of sample S1-S4 along $\phi$ = 0$^{\circ}$ (EA), 30$^{\circ}$, 60$^{\circ}$ and 90$^{\circ}$ (HA) recorded in LMOKE based microscopy at room temperature.}
	\label{fig:fig3}
\end{figure*}
that anisotropic samples give high energy product value $(BH)_{max}$ than the isotropic samples~\cite{Zhang2005}. This is because sample with magnetocrystalline anisotropy gives high
coercivity with square shaped loop, and thus, high $(BH)_{max}$ value. Although the thickness of sample S1 is twice of S2, we found similar coercivity and different anisotropy field values in samples S1 and S2. We found the enhancement of coercivity in magnetic bilayers than
the reference single magnetic layers. Interfacial exchange coupling might be a reason for this enhancement of coercivity. Also, we found the tuning of coercivity by alternating the order of magnetic layers. This indicates the presence of different interfacial exchange coupling strength in the samples S3 and S4. The magnetization reversal behaviour of samples S3 and S4 is like a rigid magnetic system because the soft and hard phases reverse with a single coercive field.

Magnetic domain images of samples S1 ((a)-(d)), S2 ((e)-(h)), S3 ((i)-(l)) and S4 ((m)-(p)) along $\phi$ = 0$^{\circ}$ (EA), 30$^{\circ}$, 60$^{\circ}$ and 90$^{\circ}$ (HA) are shown in figure 7. We found big branch domains along EA for all the samples. Further, magnetization reversal is occuring via 180$^{\circ}$ domain wall motion.  We found the nucleation and propagation of domain walls in all the samples. Magnetization reversal of samples S1, S2 and S4 away from EA occurs via big domains indicating the presence of anisotropy inhomogeneity ~\cite{Mallick2018}. However, magnetization reversal of sample S3 occurs via small domains indicating strong uniaxial anisotropy in this sample. The absence of magnetic domains is found along HA in all the samples, thus, the magnetization reversal occurs via coherent rotation. 

\subsection{FMR analysis}
In order to understand the anisotropy symmetry, we have performed in-plane angle ($\phi$) dependent FMR measurements at an interval of 10$^{\circ}$.

We can write the magnetic free energy density as the equation given below~\cite{Mallick2018,Mallik2018b}; 
\begin{eqnarray}
 E = HM_{S}[\sin\theta_{H}\sin\theta\cos(\phi-\phi_{H})+\cos\theta_{H}\cos\theta] \nonumber\\
 -2\pi(M_{S})^2(\sin\theta)^2+K_{P}(\sin\theta)^2 \nonumber\\
 +K_{in}(\sin\theta)^2(\sin(\phi-\phi_{0})^2
\end{eqnarray}
\begin{figure*}[hbt!]
	\centering
	\includegraphics[height=4cm]{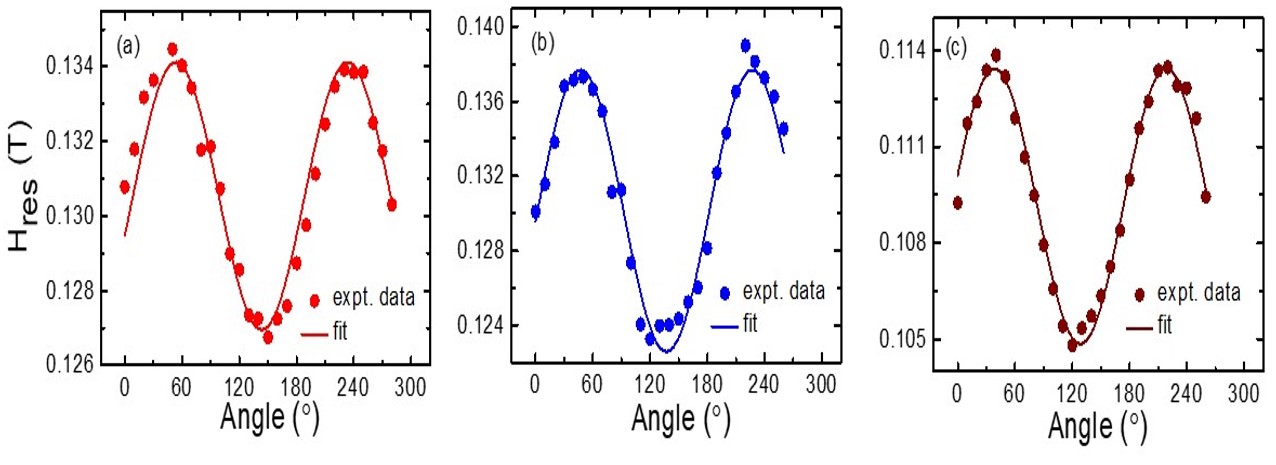}
	\caption{(a)-(c): The plot of resonance magnetic field ($H_{res}$) vs in-plane angle $\phi$ for samples S1, S3 and S4, respectively. Solid symbols are the experimental data while solid lines are the best fit using eq. 2.}
	\label{fig:fig5}
\end{figure*}
where, perpendicular uniaxial anisotropy and in-plane two-fold uniaxial anisotropy constants are defined as $K_{P}$ and $K_{in}$, respectively. The angles of applied magnetic field $\it H$ and saturation magnetization $M_{S}$ wrt z-axis are denoted as $\theta_{H}$ and $\theta$, respectively. $\phi_{H}$ is the angle of projection of $M_{S}$ in x-y plane wrt x-axis. $\phi$ is the angle of the projection of $\it H$ in the x-y plane wrt x-axis. $\phi_{0}$ is the two-fold EA direction wrt the x-axis. The directions of $M_{S}$, $\it {H}$ and the two fold EA $\phi_{0}$ can be found in our previous work by Mallick et al.~\cite{Mallick2018}.

It should be noted that the magnetic field was applied in the film plane. Therefore we have used the following dispersion relation to fit the angle dependent $H_{res}$ in order to find the values of $H_{K}$ and $h_{u}$~\cite{Mallick2018}. 

\begin{eqnarray}
(\frac{\omega}{\gamma})^2=[H_{res}\cos(\phi-\phi_{H})-h_{U} \nonumber\\
+H_{K}(\sin(\phi-\phi_{0}))^2][H_{res}\cos(\phi-\phi_{H}) \nonumber\\
+H_{K}+2H_{K}(\sin(\phi-\phi_{0}))^2] 
\end{eqnarray}

where, $h_{u}$ = $\frac{2K_{P}}{M_{S}}$ - 4$\pi$$M_{S}$ and $H_{K}$ = $\frac{2K_{in}}{M_{S}}$.
In-plane angle dependent FMR measurements are performed at a fixed frequency of 9 GHz. FMR measurement confirms the presence of uniaxial magnetic anisotropy in all the samples. Figure 8 shows the plot of in-plane angle dependent $H_{res}$. The solid scattered data points are the experimental data whereas the solid continuous line is the fitted data using eq. 2. We could not find the ferromagnetic resonance signal of sample S2, therefore the plot of $H_{res}$ vs $\phi$ of this sample has not been shown. $H_{K}$ values of 0.0036 T, 0.0082 T and 0.0041 T are evaluated for samples S1, S3 and S4, respectively, by fitting the experimental data (fig. 8) using eq. 2.
\begin{figure*}[hbt!]
	\centering
	\includegraphics[height=5cm]{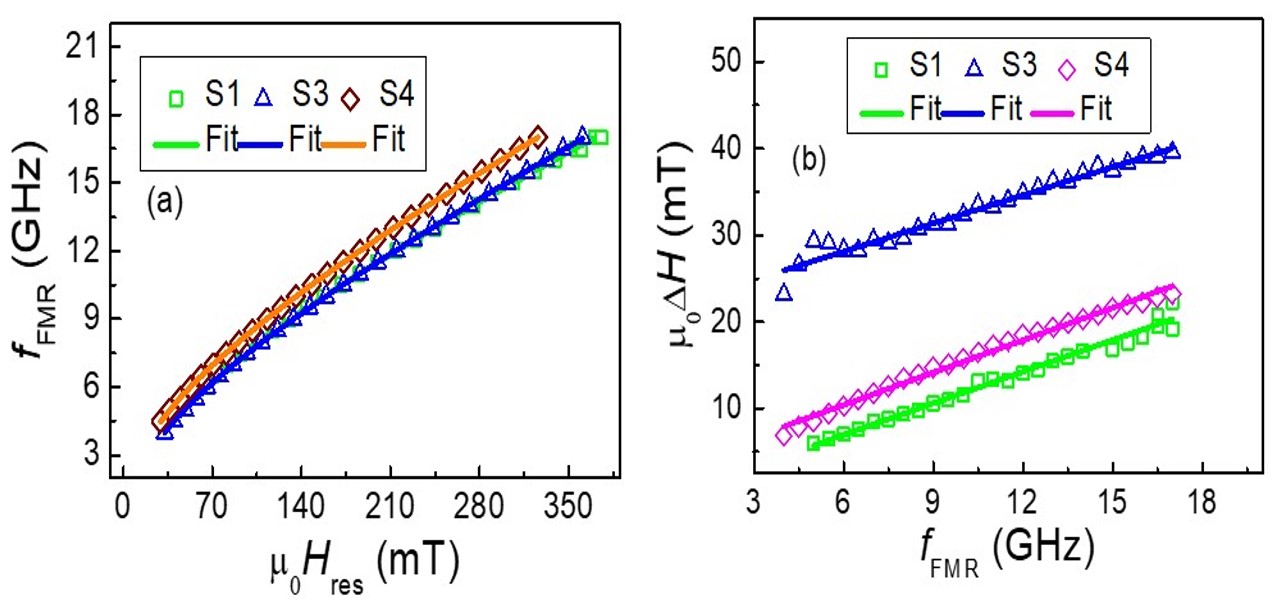}
	\caption{(a) $H_{res}$, (b) $\Delta H$ versus $f_{FMR}$ plot and their fits using eqs. (3) and (4) for the samples S1, S3 and S4.}
	\label{fig:fig6}
\end{figure*}
Figure 9 shows the FMR frequency ($f_{FMR}$) vs $H_{res}$ and line width ($\Delta H$), respectively. The effective demagnetization field (4$\pi$$M_{eff}$), effective anisotropy field ($H_{Keff}$) and the gyromagnetic ratio $\gamma$ = $\frac{g\mu_{B}}{\hbar}$ values have been extracted by fitting experimental data (fig. 9 (a)) using the following Kittel equation~\cite{Kittel1948,Singh2017}: 
\begin{eqnarray}
f_{FMR} &=\frac{\gamma}{2\pi} \sqrt{(4\pi M_{\rm eff} + H_{\rm res} + H_{\rm Keff})(H_{\rm res} + H_{\rm Keff})}
\end{eqnarray}

Similarly, the Gilbert damping constant value $\alpha$ is obtained by fitting the line width ($\Delta H$) vs $f_{FMR}$ (fig. 9 (b)) using the following equation~\cite{Singh2017,Kalarickal2006};

\begin{eqnarray}
\Delta H = \Delta H_{\rm 0} + \frac{4 \pi \alpha f_{\rm FMR}}{\gamma}  
\end{eqnarray}

where, $\Delta H_{0}$ is the inhomogeneous linewidth broadening.

Due to large linewidth broadening, we could not measure the FMR spectra of sample S2, and hence, the $H_{res}$ and $\Delta H$ values. It is theoretically reported that line width value depends on anisotropy field $H_{K}$ and the interlayer exchange coupling of two ferromagnetic layers separated by a non-magnetic layer~\cite{Layadi2015}. We found different values of $H_{K}$ in the bilayer samples S3 and S4 from Kerr microscopy measurements. This result indicates different interfacial exchange coupling strength from the ferromagnetic bilayers. The large increase in the linewidth value of sample S3 in comparison to S4 may be due to the change in interfacial exchange coupling of the bilayers. Also, little deviation in the $H_{res}$ value from all other samples is observed in sample S4 and this may be due to modified exchange coupling at the interface of the ferromagnetic layers. We reported earlier that direct exchange coupling between magnetic layer leads to the enhancemet of the Gilbert damping constant $\alpha$ value~\cite{Nayak2019}. We are getting simultaneously high coercivity and less $\alpha$ values in sample S3 which is good for FMR applications. Omelchenko et al., reported the tuning of damping by alternating the order of Py/Fe bilayers deposited on Si substrate with Ta as seed layer~\cite{Omelchenko2017}.  However, in this study, damping remains similar by alternating the order of magnetic layers which is useful for potential applications. Table 5 shows the list of values of g, 4$\pi$$M_{eff}$ , $H_{Keff}$, $\alpha$, $\Delta H_{0}$, and $K_{S}$ of all the samples. 

\begin{table*}[hbt!]
	\caption {List of values of the magnetic parameters g, 4$\pi$$M_{eff}$ , $H_{Keff}$, $\alpha$, $\Delta$$H_{0}$, and $K_{S}$ for all the samples.}
	\footnotesize
	\begin{tabular*}{\textwidth}{@{\extracolsep{\fill}}lllllll}
		\hline
		Sample name&g&$\mu_{0}$4$\pi$$M_{eff}$ (mT)&$\mu_{0}$$H_{Keff}$ (mT)&$\alpha$&$\mu_{0}$$\Delta$$H_{0}$ (mT)&$K_{S}$ ($erg.cm^{-2}$) \\
		\hline
		S1 & 1.956$\pm$ 0.007 & 636.25$\pm$ 8.69 & 8.44 $\pm$0.34 & 0.0160$\pm$ 0.0005 & 0.31$\pm$0.41 & -0.042$\pm$ 0.002 \\
		S3 & 2.032$\pm$ 0.016 & 630.47$\pm$  15.83 & 2.35$\pm$0.47 & 0.0150$\pm$ 0.0006 & 21.59$\pm$0.48 & -0.148$\pm$ 0.007 \\
		S4 & 2.060$\pm$ 0.002 & 731.63$\pm$ 2.73 & 2.43$\pm$0.05 & 0.0180$\pm$ 0.0003 & 2.92$\pm$0.28 & -0.025$\pm$ 0.001 \\
		\hline
	\end{tabular*}\\
\end{table*}

In a crystalline material, due to symmetry in crystal lattice, average value of orbital angular momentum is zero. But, the orbital contribution of magnetic moment $\mu_{L}$ is non-zero leading to the g-factor greater than 2 following the relation $\it{g}$ $\simeq$ 2 (1 + ({$\mu_{L}$}/{$\mu_{S}$})) ~\cite{Nibarger2003}. As the surfaces and interfaces break inversion symmetry, that leads to crystal field no longer symmetric. Therefore, g-factor is less than 2 and follows the relation $\it{g}$ $\simeq$ 2 (1 - ({$\mu_{L}$}/{$\mu_{S}$}))~\cite{Nibarger2003}. We observed large value of the inhomogeneous linewidth broadening $\Delta H_{0}$ in sample S3 and this value is higher than sample S4.
We evaluated the volume anisotropy field $H_{K}$ for all the samples using Kerr microscopy measurements. However, using FMR spectroscopy, we can evaluate surface induced anisotropy known as perpendicular surface anisotropy constant $K_{S}$.  The effective demagnetization field (4$\pi$$M_{eff}$) and saturation magnetization $M_{S}$ values follow the below relation;

\begin{eqnarray}
4\pi M_{eff} = 4\pi M_{S}+\frac{2 K_{S}}{M_{S} t_{FM}}
\end{eqnarray}

We found $M_{S}$ value of 762 $emu.cc^{-1}$ in sample S3 which is higher than sample S1 (639 $emu.cc^{-1}$). Thus, direct exchange coupling between the magnetic layers results higher value of $M_{S}$. Also, we found $M_{S}$ value of 860 $emu.cc^{-1}$ in sample S2. Again, samples S3 and S4 ( 636 $emu.cc^{-1}$) have dissimilar $M_{S}$ values indicating the tailoring of the interfacial exchange coupling by alternating the order of magnetic layers. 

Therefore it is observed that in sample S3 i.e. when NiFe layer is grown on top of Fe, the sample exhibits high coercive field and anisotropy. Further this sample also exhibits damping value comparable to the reference single NiFe film.

\section{Conclusions:} 
We have studied the role of interface modification on the magnetization reversal of the Fe/NiFe bilayer system fabricated by magnetron sputtering. Kerr Microscopy data showed single-phase hysteresis loops, indicating strong interfacial exchange coupling. Quantitative analysis of the Kerr loops revealed the enhancement of the magnetic parameters such as coercive field  $H_{C}$ and anisotropy field $H_{K}$ which get almost doubled when NiFe layer grows over the Fe layer. Further, this bilayer sample showed smaller domains away from the EA, confirming the presence of high uniaxial magnetic anisotropy in it. The presence of uniaxial magnetic anisotropy was also revealed from the in-plane angle-dependent FMR study. By comparing the PNR results of the bilayer samples, we observed the modification of the Fe-NiFe interfacial layer upon changing the order of the magnetic layers. The strength of the interfacial exchange coupling was higher when the NiFe layer is grown over the Fe layer. Despite different values of the anisotropy field and modified interfacial exchange coupling, the Gilbert damping constant of the bilayer systems remains similar to single NiFe layer. In summary, interchanging the order of magnetic layers plays a key role in tuning the interfacial exchange coupling through modification of interdiffusion layer thickness and magnetic moment. In this respect PNR has been proven to be an ideal technique to reveal the interface magnetic properties. Tuning of fundamental magnetic properties is possible by this methodology whereas the Gilbert damping constant remains similar which is good for applications. 
\section*{Acknowledgments:}
We acknowledge the financial support from Department of Atomic Energy (DAE), Department of Science and Technology- Science and Engineering Research Board (SB/S2/CMP-107/2013) and Department of Science and Technology (DST) Nanomission (SR/NM/NS-1018/2016(G)), Government of India for providing the financial support. SB and SSD acknowledge the financial support from Newton Fund to carry out the PNR experiments at Rutherford Appleton Laboratory. BBS acknowledges DST for INSPIRE faculty fellowship. ISIS DOI for the data set along with the RB number of the experiment is 10.5286/ISIS.E.RB1520249.

\end{document}